\documentclass[a4paper,twocolumn,8pt]{revtex4}
\usepackage{graphicx}
\usepackage{fancyhdr}
\usepackage{amsmath,amsfonts,amssymb}
\usepackage{enumitem}

\usepackage{pgf,tikz}
\usetikzlibrary{arrows}

\usepackage[a4paper]{geometry}
\usepackage{color}
\definecolor{red}{rgb}{1,0,0}
\definecolor{gre}{rgb}{0,1,0}
\definecolor{blu}{rgb}{0.05, 0.11, 0.50}
\definecolor{white}{rgb}{1,1,1}

\newcommand{\sk}{\vskip0.3truecm}

\newcommand{\cc}[1]{}

\usepackage{hyperref}   
\hypersetup{
	bookmarks=true,           
	unicode=false,          
	pdftoolbar=true,        
	pdfmenubar=true,        
	pdffitwindow=false,     
	pdfstartview={FitH},    
	pdftitle={Jaynes \& Shannon's Constrained Ignorance on Entropy},    
	pdfauthor={Thomas Cailleteau aka Thomaths Quantix, Aur\'elien Barrau},     
	pdfsubject={Entropy, Ignorance, Jaynes and Shannon},   
	pdfcreator={Thomas Cailleteau, Aur\'elien Barrau},   
	pdfproducer={Producer}, 
	pdfkeywords={entropy} {ignorance} {jaynes} {Shannon}, 
	pdfnewwindow=true,      
	breaklinks=true,       
	urlcolor= blu,        
	linkcolor= red,        
	colorlinks=true,       
	linkcolor=blu, 
	citecolor=blu,        
	filecolor=blu,      
	urlcolor=blue           
}

\begin{document}

\title{Jaynes \& Shannon's Constrained Ignorance and Surprise}
	
\author{Thomas Cailleteau}

\affiliation{
Sant Job Skolaj-Lise,  42 Kerguestenen Straed, 56100 BroAnOriant, Breizh}
\date{\today}

\email{thomas.cailleteau@lpsc.in2p3.fr}
	
	\begin{abstract}
In this simple article, with possible applications in theoretical and applied physics, we suggest an original way to derive the expression of Shannon's entropy from a purely variational approach, using constraints. Based on the work of Edwin T. Jaynes, our results are not fundamentally new but the context in which they are derived might, however, lead to a remarkably consistent formalism, where the maximum entropy principle appears naturally. After having given a general definition of ``ignorance" in this framework, we derive the somehow general expected expression for the entropy using two approaches. In the first, one is biased and has a vague idea of the shape of the entropy function. In the second, we consider the general case, where nothing is {\it a priori} known. The merits of both ways of thinking are compared.
	\end{abstract}
	
\maketitle
\tableofcontents

\section{Introduction}

This work, grounded in Edwin T. Jaynes' book \textit{Probability Theory: The Logic of Science} \cite{Jaynesbook}, could be useful both for formal or practical purposes \cite{Aspnesetall}.
In information theory, the entropy of a random variable is the average level of ``information", ``surprise", or ``uncertainty" associated with the possible possible outcomes of the considered variable. It was first introduced by Claude Shannon in 1948 \cite{wikishannon1} and shares its name \footnote{As pointed out in \cite{Tribus}, it seems that it was actually von Neumann who told Shannon to call the function he was studying ``entropy" as its characteristics were close to those of the usual entropy. \textit{``My greatest concern was what to call it. I thought of calling it ``information", but the word was overly used, so I decided to call it ``uncertainty".  When I discussed it with John von Neumann, he had a better idea. Von Neumann told me, ``You should call it entropy, for two reasons. In the first place your uncertainty function has been used in statistical mechanics under that name, so it already has a name. In the second place, and more importantly, nobody knows what entropy really is, so in a debate you will always have the advantage"."} } with the entropy used in thermodynamics and statistical physics while no rigorous formal correspondence between both of them haw been strictly demonstrated. 

In the following, we investigate how it is possible to recover the expression of the entropy from a low level approach, with few assumptions about the context, in the spirit of \cite{Jaynesbook}. We also consider the axiomatic construction of the notion of ``surprise" \cite{bookRoss} and comment on this. We define a quantity we call \textit{ignorance} instead of \textit{incertitude} as it seems to fit better with the constraints used in this framework. Imposing that it should be continuous, symmetrical, and should keep its structure in any sub-situation, we derive \footnote{To the best of our knowledge this has never been done in this precise way but we would be glad to receive any comment if we are wrong.} results leading to a clear expression for the entropy. We investigate some technical subtleties  expressing our (real) ``ignorance" to avoid biases in the calculations. The resulting formalism seems appealing and might lead to some deeper insights on this question.  	


	\section{Ignorance}

Let us consider a variable $x$ which can take on $n$ different discrete values $(x_1, .., x_n)$  corresponding to $n$ different propositions $(A_1, .., A_n)$.  The basic question is:
\begin{flushleft}
	\textit{What probabilities $(p_1, .., p_n)$ should we assign to the possibilities $(x_1, .., x_n)$ ? }.
\end{flushleft}

\subsection{What are the available knowledges? }

\begin{itemize}
	\item[$\bullet$] The sum of all probabilities is equal to one,
	\begin{equation}\label{eq:totalprobas}
	\displaystyle \sum_{i=1}^n p_i = 1, 
	\end{equation}
	therefore, the \textit{"ignorance of the knowns"} associated to this information is simply 0 and could be expressed  as		
	\begin{equation}\label{eq:ignoranceprobasomme}
	H[\lambda,p_1,..,p_n] = \lambda_0(x) \left(\sum_{i = 1}^{n} p_i - 1  \right),
	\end{equation}		
	where $h[p_1, ..,p_n] =\displaystyle \left(\sum_{i = 1}^{n} p_i - 1  \right)$ is a constraint obtained after derivation with respect to $\lambda_0(x)$, a general Lagrange multiplier. 		
	\begin{enumerate}
		\item At this stage, $x$ is just a set of yet-to-be-determined variables. In this work we consider the Lagrange multiplier te be constant or, a least, to be independent of the probabilities $p_i$. However, in principle, it might be interesting to also consider other situations which could allow to use the formalism beyond the maximization of entropy issue.
		\item The expression of the ignorance of the knowns given by Eq.(\ref{eq:ignoranceprobasomme}) takes a simple form. However, in some circumstances, one might consider a more general expression like 		
		\begin{equation}\label{eq:ignoranceprobasomme_bis}
		H[\lambda,p_1,..,p_n] = \dfrac{1}{m} \lambda_0(x) \left(\sum_{i = 1}^{n} p_i - 1  \right)^m,
		\end{equation}	
		for all $m \in \mathbb{N}^*$ (it has to be positive to prevent any divergence after dividing by the constraint). The factor $\dfrac{1}{m}$ avoids the need for a rescaling after the derivation. As will be explained later, one could, in principle, perform the calculation and rescale it by the infinity factor (expressed for instance in the term $ln(x+y-1)$ as $x+y \rightarrow 1$). But whatever the choice of $m>0$, due to constraint, this ignorance will always give 0 in the final expression and we expect that, in this formalism, this will change nothing to the result: two robots -- to refer to the usual image -- carrying out the same calculations with different values of $m$ are expected to derive the same result for the expression of the ignorance/probabilities. This will be confirmed at the end, together with some statements on the preferred settings.
	\end{enumerate}
	
	\item[$\bullet$] Let now assume that we have another knowledge taking the form of a set of $k$ constraints about the probabilities, $k \leq n$, 		
	\begin{equation}\label{eq:setofconstraints}
	f_i[p_1,...p_n] = 0, \hskip0.5truecm \forall 0<i\leq k,
	\end{equation}
	the associated ignorance, also vanishing, would be as previously:
	\begin{equation}\label{eq:ignorancebis}
	\displaystyle H[\lambda_i,p_1, .., p_n] = \sum_{i=1}^k \lambda_i(x) f_i[p_1,...p_n].
	\end{equation}
	For instance, it could be that then $p_1 = 2p_2$. What are the consequences in this formalism ? 
	
\end{itemize}

\subsection{Requirements on the ignorance} 

So far we have dealt with known notions, leading to a vanishing ignorance. However, we want to consider ignorance in its literal sense, that is \textit{"lack of knowledge or information"}. The function $H$, as defined before, is a way to assign a "degree" about the global situation. The requirement should be:
\begin{enumerate}
	\item \textbf{Continuity} : $H$ has to be continuous, so that changing the values of the probabilities by a very small amount should only change the ignorance by a small amount.
	\item \textbf{Symmetry} : $H$ has to be unchanged if the outcomes $p_i$ are re-ordered.
\end{enumerate}

As made clear by the original work of Shannon on the derivation of the entropy, one could think about ignorance/uncertainties as the total \textit{expected}/average  ignorance, having put all the information we know at the beginning of the calculations. In \cite{Jaynesbook}, 
Jaynes argues that we should carry out, at some point, a ``variational approach". This work is a simple attempt in this direction.  

One might expect the total ignorance to be, in this framework, such that 
\begin{equation}
H_{tot} = H_{knowns} + H_{unknowns}.
\end{equation}
whith $H_{knowns} = 0$ and $H_{unknowns} = \sum_i p_i H_i$, the average of the ignorances. However, at the end, we will relax this expression, rather setting only $H_{unknowns} = \sum_i H_i$.

\subsubsection{At first, $H_{unknowns} = \sum_i p_i H_i$}

As somehow explained by Shannon and Jaynes, let us imagine that at first the robot is aware of three propositions $(A_1,A_2,A_3)$ of unknown probabilities $p_1$, $p_2$ and $p_3$. The ignorance of the robots would therefore be 

\begin{equation}
\displaystyle H[p_1,p_2,p_3] = \dfrac{\lambda(x)}{m} \left(\sum_{i=1}^3 p_i-1\right)^m +  \sum_{i=1}^3 p_i H_i[p_i] .
\end{equation}
In the case $m=1$, it is just  
\begin{eqnarray*}
	&& H[p_1,p_2,p_3] = \\
	&& \lambda(p_1+p_2+p_3-1) + p_1 H_1[p_1] + p_2 H_2[p_2]+p_3 H_3[p_3] 
\end{eqnarray*}

\subsubsection{After an update}

Then, as illustrated below, the robot learns that the third propositions may in fact be a combination of three (or less or more) sub-propositions of probabilities $(v_1,v_2,v_3)$ with $v_1+v_2+v_3 = p_3$, with thus $ v_i = p(A'_{i} | A_3) $. The situation is represented by the tree below, which is not here a tree diagram of probabilities in the usual sense ($\sum_i v_i \neq 1$).

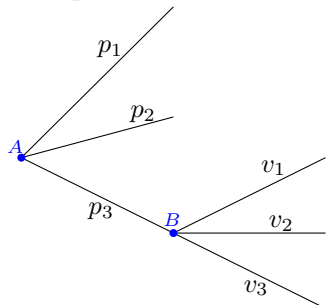
\definecolor{qqqqff}{rgb}{0,0,1}
\definecolor{cqcqcq}{rgb}{0.75,0.75,0.75}
\begin{tikzpicture}
\draw (0,0)-- (2,2);
\draw (0,0)-- (2,0.54);
\draw (0,0)-- (2,-1);
\draw (2,-1)-- (4,0);
\draw (2,-1)-- (4,-1);
\draw (2,-1)-- (4,-2);
\draw (0.88,1.67) node[anchor=north west] {$p_1$};
\draw (1.31,0.81) node[anchor=north west] {$p_2$};
\draw (0.76,-0.47) node[anchor=north west] {$p_3$};
\draw (3.03,0.09) node[anchor=north west] {$v_1$};
\draw (3.13,-0.64) node[anchor=north west] {$v_2$};
\draw (2.8,-1.5) node[anchor=north west] {$v_3$};
\begin{scriptsize}
\fill [color=qqqqff] (0,0) circle (1.5pt);
\draw[color=qqqqff] (-0.08,0.15) node {$A$};
\fill [color=qqqqff] (2,-1) circle (1.5pt);
\draw[color=qqqqff] (1.98,-0.82) node {$B$};
\end{scriptsize}
\end{tikzpicture}

The "sub-ignorance" for the proposition $A_3$ would therefore be written 
\begin{equation}\label{eq:subignorance_B_before}
p_3 \times H_3[p_3] = \dfrac{\mu(x)}{m} \left( \sum_{i=1}^3 v_i -p_3 \right)^m + \sum_{i=1}^3 v_i H_i[v_i],
\end{equation}
leading to an update of the previous ignorance, 
\begin{eqnarray}
&& H[p_1,p_2,p_3] = \\ 
&&\lambda(p_1+p_2+p_3-1) + p_1 H_1[p_1] + p_2 H_2[p_2] \nonumber\\
&& + \mu(v_1+v_2+v_3 - p_3) + v_1 H'_1[v_1] +v_2 H'_2[v_2] +v_2 H'_2[v_2].\nonumber
\end{eqnarray}
In this case we are dealing with another constraint $\mu(x)$, illustrating what said previously for the ignorance in Eq.(\ref{eq:ignorancebis}). \sk 

After the update, the robot is having now five propositions $A_i$, $1\leq i \leq 5$ in total, of propailities $p_i$, and so has an updated expected ignorance
\begin{equation}\label{eq:sommetotalignorance}
\displaystyle H[p_1,p_2,p_3,p_4,p_5] = \dfrac{\lambda(x)}{n} \left( \sum_{i=1}^5 p_i -1 \right)^n + \sum_{i=1}^5 p_i H_i[p_i].
\end{equation}

\subsubsection{Remarks}

Moreover, after dividing the ignorance in Eq.(\ref{eq:subignorance_B_before}) by $p_3$, setting $w_i = \dfrac{v_i}{p_3}$, therefore $\displaystyle \sum_{i} \dfrac{v_i}{p_3} = \sum_i w_i = 1$, and rescaling the Lagrange multiplier $\mu(x) \rightarrow  \dfrac{\mu(x)}{p_3}$, one "gets back" probabilities such that
\begin{equation} \label{eq:subignorance_B}
H_3\left[\dfrac{v_1}{p_3},\dfrac{v_2}{p_3},\dfrac{v_3}{p_3}\right] = \dfrac{\mu(x)}{m} \left( \sum_{i=1}^3 w_i -1 \right)^m +\sum_{i=1}^3 w_i H_i[w_i],
\end{equation}
where $p_3$ is considered here as a constant parameter. Ignorance in Eq.(\ref{eq:subignorance_B}) is simply the one the robot would have if it does not know about the previous propositions other than $A_3$, thus its state of knowledge starting at the node B is: \sk

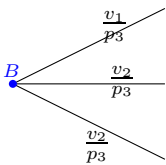
\definecolor{qqqqff}{rgb}{0,0,1}
\definecolor{cqcqcq}{rgb}{0.75,0.75,0.75}
\begin{tikzpicture}
\draw (2,-1)-- (4,0);
\draw (2,-1)-- (4,-1);
\draw (2,-1)-- (4,-2);
\draw (3.03,0.09) node[anchor=north west] {$\frac{v_1}{p_3}$};
\draw (3.13,-0.64) node[anchor=north west] {$\frac{v_2}{p_3}$};
\draw (2.8,-1.5) node[anchor=north west] {$\frac{v_2}{p_3}$};
\begin{scriptsize}
\fill [color=qqqqff] (2,-1) circle (1.5pt);
\draw[color=qqqqff] (1.98,-0.82) node {$B$};
\end{scriptsize}
\end{tikzpicture}

\subsubsection{In a nutshell}

Taking into account the possible updates we yet do not know, the general expression of the ``Ignorance" we are dealing with so far is therefore, as we are biased, 
\begin{equation}\label{eq:totalignorance}
H_{tot} [p_1,..,p_n] = H[\lambda_{\mu},p_1,..,p_n] + \displaystyle \sum_{i=1}^{n} p_i H_i \left[\dfrac{v_1}{p_i}, .., \dfrac{v_r}{p_i}\right].
\end{equation}
In the same way, we could generally has set
\begin{equation}\label{eq:totalignorance_not_weighter}
H_{tot} [p_1,..,p_n] = H[\lambda_{\mu},p_1,..,p_n] + \displaystyle \sum_{i=1}^{n} H_i \left[\dfrac{v_1}{p_i}, .., \dfrac{v_r}{p_i}\right],
\end{equation}
with, at least in this case,
\begin{equation}\label{eq:constraint_n_dependent}
\displaystyle H[\lambda_\mu,p_1,..,p_n] = \dfrac{1}{m} \lambda(u) \left( \sum_{i=1}^n p_i -1   \right)^m = 0
\end{equation} 
is the constraint dealing with the information that we know -- associated with no ignorance -- and $H_i [p_i]$ the \textit{local} ignorance about the situation on each branch.

Consequently, we could ask \textit{"what are the probabilities that minimize/maximize our ignorance ?"} which, at first, would lead us to an expression for the ignorance which should be at least similar to the one of the information entropy, and then, to the (usual) expression of the probabilities. Once again, we simply face the maximum entropy principle.\\

\section{Extremization of the ignorance}

In the following, we study the case where we express the ignorance for different values of $m$, first as a training and then in the general case.

\subsection{Knowing part of the expression of $H$}

From Eq.(\ref{eq:constraint_n_dependent}), we consider the constraint
\begin{equation}
\displaystyle H[\lambda_u,p_1,..,p_n] = \lambda(u)\dfrac{1}{m} \left( \sum_{i=1}^n p_i -1   \right)^m = 0
\end{equation} 
for $m\geq 1$, whose variation with respect to $p_i$ gives simply $\delta H[\lambda_u,p_1,..,p_n] = \displaystyle \sum_{w = p_i}\delta w \lambda(u) \left(\sum_{i=1}^n  p_i -1 \right)^{m-1}.\!\!\!\!\!\!$ 

The variation of the total ignorance leads to 
\begin{eqnarray}
\delta H &=&  \displaystyle \sum_{w = p_i} \delta w \left[ \lambda(u) \left(\sum_{i=1}^n  p_i -1 \right)^{m-1}\!\!\!\!\!\!+ H_w\left(x_1, .., x_n\right) \right. \label{eq:deltaHstep5} \\
&& \left.-  \sum_{i=1}^m  x_i \times \dfrac{\partial H_w}{\partial x_i}\left(x_1, .., x_n\right)  \right], \nonumber 
\end{eqnarray}
where $v_i$ has been extracted from $H_w$ via $x_i= \dfrac{v_i}{p_j}$, as it will be clear in the following. However, we know that $\displaystyle \sum_j x_j = 1$, therefore this information should appear at some point. 
Will it change something ? To see it, we will set this information on the $x_i$ at two places, with factor $p$ and $q$ taken values in $\{0;1\}$ in Eq.(\ref{eq:totalignorance}) and consider the sub-propositions of probabilities $x_i $

\subsubsection{The general equations}

Considering the expression
\begin{equation}
H_{tot} [p_1,..,p_n] = H[\lambda_{\mu},p_1,..,p_n] + \displaystyle \sum_{i=1}^{n} p_i H_i \left[\dfrac{v_1}{p_i}, .., \dfrac{v_r}{p_i}\right],
\end{equation}
from the reasoning in Eq.(\ref{eq:subignorance_B}), we could set
\begin{eqnarray} \label{eq:redundancy_eee}
H_i\left[x_1, .., x_r \right] &=& \dfrac{\mu(u)}{b} \left( \sum_{j=1}^r x_j -1 \right)^b +\sum_{j=1}^r x_j H_j[x_j] \\
&=& \dfrac{\mu(u)}{b} \left( \sum_{j=1}^r x_j -1 \right)^b + f[x_j] \label{eq:redundancy_eee_2}
\end{eqnarray}
as the unknown variable in our calculations is the $H_k$. Moreover, we could have continued and express again the sub-ignorance $H_j[x_j]$ further, in terms of the sub-sub-ignorance, but it would have been redundant as we would process to the same calculations at each node of the probability tree, again and again. Therefore here, our unknown variable is simply $f[x_j]$ which represents the situation from the probability tree of nodes A and $B_i$ for each $p_i$, shown before. 


\begin{widetext}
	\begin{eqnarray}
	\delta H & = &\displaystyle \sum_{w = p_i} \delta w  \left[  \lambda(u) \left( \sum_{i=1}^n p_i -1 \right)^{m-1}\!\!\!\!\!\! + H_w\left[\dfrac{v_1}{w},\dfrac{v_2}{w}, ..., \dfrac{v_m}{w}\right] + w \times  \sum_{i=1}^m  \dfrac{\partial \left(\dfrac{v_i}{w} \right)}{\partial w} \times \dfrac{\partial H_w}{\partial \left(\dfrac{v_i}{w} \right)}\left[\dfrac{v_1}{w},\dfrac{v_2}{w}, ..., \dfrac{v_m}{w}\right]  \right] \\
	&=& \displaystyle \sum_{w = p_i} \delta w  \left[  \lambda(u) \left( \sum_{i=1}^n p_i -1 \right)^{m-1}\!\!\!\!\!\! + H_w\left[x_i\right]  -  \sum_{i=1}^m  x_i \times \dfrac{\partial H_w}{\partial x_i}\left[x_1, .., x_n\right]  \right]  \\
	&=& \displaystyle \sum_{w = p_i} \delta w  \left[  \lambda(u) \left( \sum_{i=1}^n p_i -1 \right)^{m-1}\!\!\!\!\!\! + \left( p \dfrac{\mu(u)}{b} \left( \sum_{j=1}^r x_j -1 \right)^b + f[x_j] \right) -  \sum_{i=1}^m  x_i  \dfrac{\partial}{\partial x_i}\left( q \times \dfrac{\mu(u)}{b}\left( \sum_{j=1}^r x_j -1 \right)^b + f[x_j] \right) \right]  \,\,\,\,\,\,\,\,\,\,{} \\
	%
	&\Leftrightarrow& 0 =  \lambda(u) \left( \sum_{i=1}^n p_i -1 \right)^{m-1}\!\!\!\!\!\! +p \dfrac{\mu(u)}{b} \left( \sum_{j=1}^r x_j -1 \right)^b + f[x_i]-  \left(\sum_{i=1}^r  x_i \right) \left[q\times\mu(u)\left( \sum_{j=1}^r x_j -1 \right)^{b-1} + \dfrac{\partial f[x_i]}{\partial x_i} \right] \,\,\,\,\,\, \forall   \,\, \delta w  \\
	&\Leftrightarrow& 0 =  \lambda(u) \left( \sum_{i=1}^n p_i -1 \right)^{m-1}\!\!\!\!\!\! +p \dfrac{\mu(u)}{b} \left( \sum_{j=1}^r x_j -1 \right)^b - q\times \mu(u) \left(\sum_{i=1}^r  x_i \right) \left[\left( \sum_{j=1}^r x_j -1 \right)^{b-1}\right] + f[x_i] -  \sum_{i=1}^r  x_i\dfrac{\partial f[x_i]}{\partial x_i} \\
	&\Leftrightarrow& 0 =  \lambda(u) \left( \sum_{i=1}^n p_i -1 \right)^{m-1}\!\!\!\!\!\! + \dfrac{\mu(u)}{b} \left( \sum_{j=1}^r x_j -1 \right)^{b-1} \left[ p(\sum_{j=1}^r x_j -1) - bq \left(\sum_{i=1}^r  x_i \right) \right] + f[x_i] -  \sum_{i=1}^r  x_i\dfrac{\partial f[x_i]}{\partial x_i} 
	\end{eqnarray}
	
	leading us therefore to solve in the general case
	\begin{equation}
	0 =  \lambda(u) \left( \sum_{i=1}^n p_i -1 \right)^{m-1}\!\!\!\!\!\! + \dfrac{\mu(u)}{b} \left( \sum_{j=1}^r x_j -1 \right)^{b-1} \left[ (p-bq) \left(\sum_{i=1}^r  x_i \right) -p\right] + f[x_i] -  \sum_{i=1}^r  x_i\dfrac{\partial f[x_i]}{\partial x_i} 
	\end{equation}

\end{widetext}

Commentaries and assumptions at this point : 
\begin{itemize}
	\item we put $p$ and $q$ in order to distinguish from where the information that $\displaystyle \sum_{i=1}^{r} x_j-1 = 0$ comes from : as it appears twice in the calculation, it may be redundant to do so and we may consider that one of the two terms could be superfluous. However, it could also play a major role in the expression of the solution when derived, and therefore we keep the $q$ in front, as such.
	\item $b=m$ : not really an assumption as by redundancy it has to be true [assuming that Ignorance at each node has the same expression]
	
	\item For simplicity, we are looking only at two sub-propositions of possibilities $x$ and $y$ such that $x+y=1$. We guess that any sub-situation can be seen as : $A_x$ "something happens", $A_y$ "something does not", and by recurrence at each node it should be true. For instance, from proposition $A_3$ we could have sub-proposition $A_{3,1} = A_3$ of probability $x=1$. 
	
	\item Regarding $\alpha(p_i, m) =  \lambda(u) \left( \sum_{i=1}^n p_i -1 \right)^{m-1} \,\, {} $, it should be "just" a constant for $f[x_j]$, that is, $x_i$ are considered now as independent of $p_j$. \cc{I think it's true but not totally sure ... [maybe it exists a clearer way to express this] }
\end{itemize}

Therefore, in the following, we will try to solve
\begin{eqnarray}\label{eq:equadiffpourf_1}
0 = && \alpha(m) + \dfrac{\mu}{b} \left( x+y -1 \right)^{b-1} \left[ (p-bq) (x+y) -p\right] + f[x_i] \nonumber \\
&& -  x \dfrac{\partial f[x,y]}{\partial x}-  y \dfrac{\partial f[x,y]}{\partial y}
\end{eqnarray}

This equation is a linear first order PDE we can rewrite as 

\begin{equation}
a(x,y) u_x + b(x,y) u_y = f(x,y,u)
\end{equation}
with $u = f(x,y)$ such that $u_z = \partial_z u = \dfrac{\partial u}{\partial z}(x,y)$ for $z=x,y$, and 
\begin{eqnarray*}
	a(x,y) &=& x, \\
	b(x,y) &=& y, \\
	f(x,y,u) &=& \alpha(m) + \dfrac{\mu}{b} \left( x+y -1 \right)^{b-1} \left[ (p-bq) (x+y) -p\right] + u
\end{eqnarray*}

Using the method of characteristics, we have to solve 
\begin{equation}
\dfrac{dx}{a} = \dfrac{dy}{b} = \dfrac{du}{f}  
\end{equation}
that is 

\begin{equation}
\dfrac{dx}{x} = \dfrac{dy}{y} = \dfrac{du}{\alpha + \dfrac{\mu}{b} \left( x+y -1 \right)^{b-1} \left[ (p-bq) (x+y) -p\right] + u} 
\end{equation}

\begin{enumerate}
	\item From the two first ones, we get 
	\begin{equation} \label{eq:caract1}
	\dfrac{dx}{x} = \dfrac{dy}{y} \Leftrightarrow y = c_1 x ,
	\end{equation}
	and so $c_1 = \dfrac{y}{x}$
	
	\item From the second ones, setting $(1+c_1)x =cx$, 
	
	\begin{eqnarray}\label{eq:caract2}
	&& \dfrac{dx}{x} = \dfrac{du}{\alpha(m) + \dfrac{\mu}{b} \left( cx  -1 \right)^{b-1} \left[ cx(p-bq) -p\right] + u} \\ 
	&\Leftrightarrow&  \dfrac{dx}{x} = \dfrac{du}{\alpha(m) + \dfrac{\mu}{b} \left( cx  -1 \right)^{b-1} \left[ (p-bq) cx -1\right] + u}	\\
	&\Leftrightarrow& \dfrac{du}{dx} = \dfrac{\alpha(m) + \dfrac{\mu}{b} \left( cx  -1 \right)^{b-1} \left[ (p-bq) cx -1\right] + u}{x} \\ 
	&\Leftrightarrow&\dfrac{du}{dx} - \dfrac{u}{x} = \dfrac{\alpha(m) + \dfrac{\mu}{b} \left( cx  -1 \right)^{b-1} \left[ (p-bq) cx -p\right]}{x}
	\end{eqnarray}
	
	\item Multiplying both side by $\dfrac{1}{x}$, we get 
	\begin{equation}
	\dfrac{1}{x} \dfrac{du}{dx} - \dfrac{u}{x^2} = \dfrac{d}{dx}\left(\dfrac{u}{x} \right) =  \dfrac{\alpha + \dfrac{\mu}{b} \left( cx  -1 \right)^{b-1} \left[ (p-bq) cx -p\right]}{x^2}
	\end{equation}

	and we have therefore to solve
	
	\begin{eqnarray}
	\dfrac{u(x,y)}{x}\!\!\!\! &= &\!\!\!\!\displaystyle \int_{a}^x  \dfrac{\alpha + \dfrac{\mu}{b} \left( c\xi  -1 \right)^{b-1} \left[ (p-bq) c\xi -p\right]}{\xi^2} d\xi + c_2\left(\dfrac{y}{x}\right) \,\,\,\,{}\,\,\,\,{}\,\,\,\,{} \\
	\!\!\!\! &= &\!\! \beta(x,y) -\dfrac{\alpha}{x} + \dfrac{\mu}{b} c \int_{ca}^{cx} \dfrac{dw}{w^2} (w-1)^{b-1}((p-bq)w-p)\,\,\,\,{}\,\,\,\,{}\,\,\,\,{}
	\end{eqnarray}	
	
	where $\beta(x,y) = c_2\left(\dfrac{y}{x} \right) + constants$ , $a$ is a constant and we set $w =c \xi$ for more simplicity.
	
\end{enumerate}

We can therefore express the "solution" as 
\begin{equation}
u(x,y) = \beta\left(\dfrac{y}{x} \right)x - \alpha(m)  + \dfrac{\mu (cx)}{b} (p I_1 - bq I_2)
\end{equation}
where
\begin{eqnarray}
I_1 &= & \displaystyle \int_{ca}^{cx} \dfrac{dw}{w^2}(w-1)^b \\
I_2 &= & \displaystyle \int_{ca}^{cx} \dfrac{dw}{w}(w-1)^{b-1} 
\end{eqnarray}

\subsubsection{What are the results of $b=1$ or $b=2$ ?}

\paragraph{case where $b=1$}${}$\\${}$

In this case, setting $m=b=1$, we have $\alpha(m) = \lambda$ the Lagrange multiplier (here considered as constant). Regarding the integrals, 

\begin{eqnarray}
I_1 &= & \displaystyle \int_{ca}^{cx} \dfrac{dw}{w^2}(w-1) = \int_{ca}^{cx} dw \left(\dfrac{1}{w} - \dfrac{1}{w^2}  \right)\\
&=& ln(cx)+ \dfrac{1}{cx} + const\\
I_2 &= & \displaystyle \int_{ca}^{cx} \dfrac{dw}{w} = ln(cx)+ const 
\end{eqnarray}	
and therefore

\begin{eqnarray}
{}\!\!\!u(x,y)& =& \beta\left(\dfrac{y}{x} \right)x - \lambda  + \mu (cx)\left(p  \times \left(ln(cx)+ \dfrac{1}{cx} \right) - q \cdot ln(cx) \right) \nonumber \\
&=& \beta\left(\dfrac{y}{x} \right)x - \lambda + \mu p  + \mu (p-q) (cx) ln(cx) \label{eq:comparauu1}
\end{eqnarray}
from which we could say that 
\begin{itemize}
	
	\item if $u(x,y) = constant \times (x+y) $, as the constraints $x+y = cx \rightarrow 1$ will be applied at the end, this will lead $u(x,y)$ to be only a constant, and we could rescale it in order to absorb it. However, the drawback of this formulation is also that .. $ln(cx) \rightarrow 1 $ as we will talk later.
	
	\item as $p =1$, then we have a $\alpha- \mu$ term, which corresponds to, as  $\alpha_{b=1} = \lambda(u)$, $\lambda(u)-\mu(u)$. Our guess would be that at each node and sub-nodes, we have the same "kind of information", and therefore we would put $\lambda(u) = \mu(u)$, leading  $\lambda(u)-\mu(u)$ to be zero. In the other way around, we would just have either to rescale by removing the constants, or either express any quantity in terms of $H[p]-H_0[P]$ where $H_0[p]$ is a reference value (the minimum, maximum, .. of the ignorance).
	
	\item if $p=q$, the logarithm term will disappear, at least for the case $b=1$. As we would like ignorance to decrease when the probabilities are known to be 0 or 1, either 
	\begin{itemize}
		\item we set $p=0$ and $q=1$, and we have with the choice of $\mu>0$ the kind of expression we need (after rescaling the expression due to terms as $\alpha$),
		\item or we set $p=1$, $q=0$ and choosing $\mu<0$ (equivalent to $\mu\left(1-\sum_i x_i\right)$ ) would give us 
		\begin{eqnarray}
		u(x) &=& \beta x + \mu (c x) ln(cx) \\
		u(x,y)	&=& \beta\left(\dfrac{y}{x} \right) x+\beta\left(\dfrac{x}{y} \right) y + \mu (x+y)ln(x+y) \label{eq:totalignorancefid}
		\end{eqnarray}
	\end{itemize}

\end{itemize}

\paragraph{case where $b=2$}${}$\\${}$

In this case

\begin{eqnarray}
I_1 &= & \displaystyle \int_{ca}^{cx} \dfrac{dw}{w^2}(w-1)^2 = \int_{ca}^{cx} dw \left(1-\dfrac{2}{w} + \dfrac{1}{w^2}  \right)\\
&=& cx - 2ln(cx) - \dfrac{1}{cx}+  const\\
&=& \dfrac{(cx-1)(cx+1)}{cx} - 2 ln(cx) \\
I_2 &= & \displaystyle \int_{ca}^{cx} \dfrac{dw}{w}(w-1) = \displaystyle \int_{ca}^{cx}dw \left(1-\dfrac{1}{w} \right) \\
&=& cx - ln(cx) 
\end{eqnarray}
and therefore, the ignorance would be
\begin{eqnarray}
u(x,y) &=& \beta x - \alpha + \dfrac{\mu(cx)}{2} \times \left[p \dfrac{(cx-1)(cx+1)}{cx}  -2q (cx) \right]  \nonumber \\
&&  +\dfrac{\mu(cx)}{2}  \times \left[-2p ln(cx) - 2q(-ln(cx) )   \right]
\end{eqnarray}
thus
\begin{eqnarray}
u(x,y) &=& \beta x - \alpha + \dfrac{\mu}{2} \times \left[p  {(cx-1)(cx+1)} -2q (cx)^2 \right]  \nonumber \\
&& - \mu (p-q) (cx) ln(cx). \label{eq:comparauu2}
\end{eqnarray}

We could say also that
\begin{itemize}
	\item regarding $- \alpha + \dfrac{\mu}{2} p  {(cx-1)(cx+1)}$, as we would expect that $m=b=2$, then 
	\begin{equation}
	\alpha(p_i) = \lambda(u)\left(\sum_{i=1}^n p_i-1\right) \sim \lambda\times (cp-1)
	\end{equation}
	both terms are constraints in $c\xi-1$ and so will vanish.
	
	\item Now, with this in mind, comparing Eq.(\ref{eq:comparauu1}) et Eq.(\ref{eq:comparauu2}), as we expect the ignorance to remain the same whatever the choice of the power $m=b$ of the constraint, we would expect no terms in $q (cx)^s$ for different values of $s$ depending on $b$. So, if this has to be true, then, we should set $q=0$ for the theory to remain coherent. However, setting $p=0$ and $q=1$ gives a $\mu (cx) ln(cx)$ term as for the case where $b=1$. However, as such, we would have to consider $\mu<0$ in order for the ignorance to behave correctly.
	
	\item The case $q=0$, $p=1$ and $\mu>0$ is of interest as it leads to the expression for the ignorance, after the constraint being applied, to be similar to Eq.(\ref{eq:totalignorancefid}), that is 
	
	\begin{equation}
	u(x,y)	= \beta\left(\dfrac{y}{x} \right) x+\beta\left(\dfrac{x}{y} \right) y - \mu (x+y)ln(x+y) \label{eq:totalignorancefid2}
	\end{equation}
	
\end{itemize}

As a consequences of the choices before, the expression for the total ignorance in Eq.(\ref{eq:totalignorance}) would be somehow

\begin{eqnarray}
&& H =\dfrac{\lambda(u)}{2} \left(\sum_{i=1}^n p_i-1 \right)^2 \\
&&+ \sum_{i=1}^n p_i \times \left[\beta \left[\dfrac{v_1}{p_i},\dfrac{v_2}{p_i}\right] + \lambda\left(\dfrac{v_1}{p_i} + \dfrac{v_2}{p_i}\right) ln\left(\dfrac{v_1}{p_i} + \dfrac{v_2}{p_i}\right) \right] \,\,\,\,\,\,{}\,\,\,{}\,\,\,{}
\end{eqnarray}

with $x=\dfrac{v_1}{p_i}$, $y=\dfrac{v_2}{p_i}$ and $v_1+v_2 =p_i$, $\mu = \lambda$. 
%
%

When we will apply it to a situation, the constraint will be fulfilled and so $H$ will reduce roughly to 
\begin{equation}
H = [0] + [1] + \lambda \sum_{i=1}^n p_i \times ln\left(\dfrac{v_1}{p_i} + \dfrac{v_2}{p_i}\right) 
\end{equation}
as for the case where $b=1$. Here $[X]$ means terms linear in $X$, and so having no consequences as the constrained are applied, and after rescaling.\sk

Commentaries : 
\begin{itemize}
	\item the case $b=2$ is appealing in the sens that for a variational problem in physics, $H$ would be similar to a Lagrangian/Hamiltonian where velocities of potential energies are globally in $\xi^2$. However, here it seems to be independent of the power, therefore this analogy is just to say.
	
	\item \textbf{More importanlty}, in order to apply the same logic at each node of the tree diagram, from Eq.(\ref{eq:subignorance_B_before}) with $v_j H_j[v_j]$,  to Eq.(\ref{eq:subignorance_B}) with $\dfrac{v_j}{p_i}H_j\left[\dfrac{v_j}{p_i}\right]$, we did a mixed-up change of variables which, even if it was logic regarding Eq.(\ref{eq:sommetotalignorance}), was also done in $H_j$. Consequently, due to the $\dfrac{1}{p_i}$ factor, differentiating with respect to $p_i$, we obtained a negative sign which leads to a logarithm solution for the ignorance (not obtained by a plus sign). But we artificially pass from $\sum_{j} v_j = p_i$ to $\sum_j x_j = 1$, \textit{i.e.} from $v_1+v_2 = p_3$ to $x + y = 1$, and so to $p_i \times ln\left(\dfrac{v_1}{p_i} + \dfrac{v_2}{p_i}\right) $ instead of $p_i \times ln(p_i)$ as expected. One way to cure it would have to look at  $H = ... + p_i H_i[p_i] = ... + v_j H_j[v_j]  \rightarrow .. +p_i \dfrac{v_j}{p_i} H_j \left[p_i\times \dfrac{v_j}{p_i}\right]$ but differentiating w.r.t $p_i$ would give much more complicated equations, and this would have been a patch to an artificially ill defined solution, as the next part shows a better way of doing it.
	
	Relately, as shown in Eq.(\ref{eq:totalignorancefid2}), we see a logarithm term which should
	\begin{enumerate}[label = \alph*)]
		\item go to zero as $x+y=1$ (except if we multiply it by $p_i$ as said just before)
		\item at this sub-node where a proposition is separated in more sub-propositions of possibilities $x$ and $y$, also give us the relation 
		\begin{equation}\label{eq:souspropentropyform}
		\beta(x,y) +  (x+y)ln(x+y)\,\,\,\,\,\,\,\, \rightarrow\,\,\,\,\,\,\,\, xln(x) + y ln(y).
		\end{equation}
		As $\beta(x,y)$ has not yet specified, we could take a specific value to remove the unwanted term, but this is again an artificial way of doing. 
	\end{enumerate}
	
\end{itemize}

As a consequence, as this first approach seems unsatisfying in our opinion, and as we expect similar expression for the entropy for all value of $m=b$, we will stop here and look at a more general and promising way at this point.

\subsection{Specifying nothing about $H_i$}\label{sec:notweightedignorance}

\subsubsection{Derivation of the solution}
Starting from the general expression
\begin{equation}\label{eq:ignorance_notpondered}
\displaystyle H[p_1, p_2, ..., p_n] = \dfrac{\lambda(x)}{m} \left( \sum_{i = 1}^{n} p_i - 1  \right)^m \!\!\!\!+ \sum_{i=1}^{n} H_i\left[\dfrac{v_1}{p_i}, .., \dfrac{v_r}{p_i}  \right]
\end{equation}
where we only require $H_i$ on the $r$ "sub"-probabilities at each sub-node for each $p_i$ (to recall, this is more coherent as each sub-tree is a probability tree, and as usual, the probabilities are multiplied from branch to branch the more we know about sub-situations, \textit{i.e.} sub-propositions). As before, 

\begin{eqnarray}
\displaystyle \delta H&=& \sum_{w= p_i} \delta w \left[ \lambda(u)\left( \sum_{i = 1}^{n} p_i - 1  \right)^{m-1} \!\!\!\!\! + \dfrac{\partial}{\partial w} \left(H_w \left[\dfrac{v_1}{w}, .., \dfrac{v_r}{w} \right] \right) \right]\,\,\,\,{}\,\,\,\,{} \\
&=&  \sum_{w= p_i} \delta w \left[ \lambda(u)\left( \sum_{i = 1}^{n} p_i - 1  \right)^{m-1} \!\!\!\!\! +  p \dfrac{\mu(u)}{b} \left( \sum_{j=1}^r \left(\dfrac{v_j}{w} \right)  -1  \right)^{b}  \right. \nonumber \\
&&{}\hskip-1truecm \left. + \sum_{j=1}^r \dfrac{\partial \left(\dfrac{v_j}{w}  \right)}{\partial w}  \dfrac{\partial}{\partial \left(\dfrac{v_j}{w}  \right)} \left(  q \dfrac{\mu(u)}{b} \left( \sum_{j=1}^r \left(\dfrac{v_j}{w} \right)  -1  \right)^{b} \!\!\!\! + H_w \left[ \dfrac{v_i}{w}  \right]   \right) \right]
\end{eqnarray}

where we put the constraints on $ \dfrac{v_j}{w}$ inside (with $q$) or outside ($p$) the derivation in order to keep it general and see how they impact the results.
\begin{eqnarray}
\displaystyle \delta H &=& \sum_{w= p_i} \delta w \left[ \lambda(u)\left( \sum_{i = 1}^{n} p_i - 1  \right)^{m-1} \!\!\!\!\! +  p \dfrac{\mu(u)}{b} \left( \sum_{j=1}^r \left(\dfrac{v_j}{w} \right)  -1  \right)^{b}  \right. \nonumber \\ 
&&{}\hskip-1truecm  \left. - \dfrac{1}{w} \sum_{j=1}^r \left( \dfrac{v_j}{w} \right) \left(  q \mu(u) \left( \sum_{j=1}^r \left(\dfrac{v_j}{w} \right)  -1  \right)^{b-1} \!\!\!\!\!\! +  \dfrac{\partial H_w[v/w ]}{\partial \left(\dfrac{v_j}{w}  \right)}   \right) \right]   \\
&=& 0\hskip0.7truecm  \Leftrightarrow \forall w \hskip0.7truecm  [ ..] = 0,
\end{eqnarray}
that is, setting $x_j = \dfrac{v_j}{w}$
\begin{eqnarray} \label{eq:equationresoudresousignorance}
&& \displaystyle \sum_{j=1}^r x_j \dfrac{\partial H[x_j]}{\partial x_j} \,\, = \,\, w \times \lambda(u)\left( \sum_{i = 1}^{n} p_i - 1  \right)^{m-1}    \\
&& + w \dfrac{\mu(u)}{b} \left( \sum_{j=1}^r x_j  -1  \right)^{b-1} \!\!\! \times \left( \left( p-\dfrac{bq}{w}  \right) \sum_{l=1}^r x_l - p  \right).\nonumber
\end{eqnarray}

Looking again at two sub-propositions $A_{w,1}$ and $A_{w,2}$, with $x = \dfrac{v_1}{w} $ and $y = \dfrac{v_2}{w} $ s.t $x+y=1$, we derive the solution.

Using for short $\alpha(p, m) =\lambda(u)\left( \sum_{i = 1}^{n} p_i - 1  \right)^{m-1}$ , and the method of characteristics as Eq.(\ref{eq:caract1}) giving $y = c_1 x \Leftrightarrow x+y =(1+c_1)x =cx$, we have to solve, as for Eq.(\ref{eq:caract2}),

\begin{eqnarray}
&&\dfrac{dx}{x} = \dfrac{du}{w\times \alpha(p,m) + w \dfrac{\mu}{b} (cx-1)^{b-1}\left( \left( p - \dfrac{bq}{w}  \right) cx - p \right)        } \\
&& \dfrac{du}{dx} = \dfrac{w\alpha}{x} + \dfrac{cw\mu}{b} \left( p - \dfrac{bq}{w}  \right) (cx-1)^{b-1} - \dfrac{w\mu p}{b} \dfrac{1}{x} (cx-1)^{b-1} \,\, \,\, \,\, \,\, \,\, \,\, {}
\end{eqnarray}
and so, $a$ being a constant, $\beta(x,y)$ having also constants (like the ones from $a$), we have 
\begin{eqnarray}
&& \displaystyle u(x,y) = c_2\left(\dfrac{y}{x}\right)  + w \times \alpha(p,m) \int_{a}^x d\xi \dfrac{1}{\xi}  \\
&& +\dfrac{cw\mu}{b} \left( p - \dfrac{bq}{w}  \right) \int_{a}^x d\xi (c\xi -1)^{b-1}  - \dfrac{w\mu p}{b}  \int_{a}^x d\xi  \dfrac{(c\xi -1)^{b-1} }{\xi} \,\, \,\, {}\,\, \,\, {}\,\, \,\, {}
\end{eqnarray}
that is, 
\begin{eqnarray} \label{eq:eqgeneralsousignorance}
&& \displaystyle u(x,y) = \beta\left(\dfrac{y}{x}\right)  + w \alpha(m,p) ln(x) +\dfrac{cw\mu}{b} \left( p - \dfrac{bq}{w}  \right) I_1 - \dfrac{w\mu p}{b}  I_2 \,\, {}\,\, \,\, {}\,\, {}\,\, \,\, {}
\end{eqnarray}
Again, if $m\geq 2$, then, as a constraint we will have $\alpha = \lambda(u) (cp-1) \rightarrow 0 $ and so this term with a logarithm vanishes when we consider the constraint in the final expression of the Ignorance. However, as the expression of $I_2$ shows, there is another logarithm term which should appear. 

\subsubsection{case where $b=1$}

A really interesting case because it is the simplest one which leads to what we expect, and even more, in what we think a coherent way. 
\begin{eqnarray}
\displaystyle I_1 \!\! &=& \!\! \int_{a}^x d\xi (c\xi -1)^{b-1} = \int_{a}^x d\xi  = [\xi]^x \rightarrow  x  \\
I_2 \!\! &=& \!\! \int_{a}^x d\xi  \dfrac{(c\xi -1)^{b-1} }{\xi}  = \int_{a}^x d\xi  \dfrac{(1 }{\xi}  = [ ln(\xi)]^x \rightarrow ln(x) \,\,\, {}
\end{eqnarray}
our solution is now 
\begin{equation}
\displaystyle u(x,y) =\beta\left(\dfrac{y}{x}\right)  + w[\alpha(m)-p\mu]ln(x) + w\mu \left( p - \dfrac{bq}{w}  \right) (cx) 
\end{equation}

Commentaries 
\begin{itemize}
	\item At the end, in the ignorance, constraints will play no major role as they do not influence it. However, we see that they appear here within the solution $u(x,y)$ via their Lagrange multiplier, and also via $cx$ for the last term. For this term in $cx$, as the constraint are satisfied when applying the solution, we have $cx\rightarrow 1$, but \underline{not} $ln(x) \rightarrow 0$ ! As a consequence, this leads to the constant $w \mu p -  \mu q $ in the expression of the ignorance. 
	\item In fact, at the end, this expression $w\mu p - \mu q$ will play no role, as it leads in Eq.(\ref{eq:ignorance_notpondered}) to the term 
	\begin{equation}
	\displaystyle \sum_{i=1}^n  (p_i \times \mu p - \mu q) = \mu p \sum_{i=1}^n p_i - \mu q \sum_{i=1}^n 1 \rightarrow \mu p - \mu q n 
	\end{equation}
	after applying the constraint and setting back $w\equiv p_i$. Giving always $n$ propositions at start, this former term is just a constant. In fact, all term linear in $w$ will be considered at the end as a constant due to the summation and the constraint. 
	
	\item Moreover, assuming that all constraints are implemented in a same way, we would set $m=b=1$, leading to $\alpha(1) = \lambda(u)$, but also that $\lambda(..) = \mu(..)$. As a consequence, the remaining term, the logarithm one, becomes $	w \lambda(..) (1-p) ln(x) $. As $x = \dfrac{v_1}{w}$, 
	\begin{eqnarray}
	&&	w \lambda(..) (1-p) ln(x)  =\lambda(..) (1-p) w (ln (v_1) - ln (w) ) \\
	&& =  (1-p) \lambda(..) ( - w\times ln(w) + w \times ln(v_1) )
	\end{eqnarray}
	
	\begin{enumerate}
		\item As we considered in our derivation that $v_j$ are independent of $w$, the last term is linear in $w$ and therefore, as for $\mu p w$, will lead to a constant in the final expression of the ignorance when constraints are applied. 
		
		Moreover, as $w = v_1 + v_2$, terms like $w \times ln(v_1)$ will have mixed terms as $(v_1 + v_2)ln(v_1)$. This is again linked to Eq.(\ref{eq:souspropentropyform}) where we encountered a similar problem, which is a consequence of the form $p \times ln(p)$.

		\item Then, dealing with the last term (except for $\beta(x,y)$ which condense the constants and help to restore the symmetry of the ignorance as $u(x,y) = u(y,x)$), we see a factor $1-p$. As $p=0$ or $1$, the only way to keep the logarithm of $w$ is to set $p=0$ : this is interesting because it was set artificially to consider the constraint on $v_i$ from outside ($v_i$ make sens only in $H_i[p_i]$), and therefore it is better as this leads to no consequences on what we expect. 
		
	\end{enumerate}
\end{itemize}	

In fact, with what we said previously, we see that whatever the value of $q \in \{ 0;1\}$, it has also no consequences on the expression of the ignorance which varies : At the end, it is like the obtained solution is given w.r.t $p_i$ but we used its consequences on sub-proposition to solve the equation w.r.t them. We could therefore have solved two equations from Eq.(\ref{eq:equationresoudresousignorance}) where $q = 0 $  or $q=1$, leading to similar solutions in $w ln(w) $ but it makes sens to consider $q=1$ as it considers the situation on the sub-node. Therefore

\begin{eqnarray}
0 &=& w \lambda - q \lambda \times (x+y) - x \dfrac{\partial H}{\partial x} - y \dfrac{\partial H}{\partial y} \\
\Leftrightarrow && 0 = \lambda(w - q(x + y)) - x \dfrac{\partial H}{\partial x} - y \dfrac{\partial H}{\partial y}   \\
\Leftrightarrow && u(x,y) = \beta\left(\dfrac{y}{x} \right) + \lambda w ln(x) - q \lambda (cx) \\
&& u(x,y) \rightarrow - 2 \lambda w\times ln(w) + [w] + [cx\rightarrow 1]
\end{eqnarray}  
as we restore the symmetry by setting $\beta\left(\dfrac{y}{x} \right) \sim \lambda w ln(y)$ and as $ln(x) + ln(y) = ln(v_1) + ln(v_2) - 2 ln(w)$. 

However, if we generalize it with more than 2 sub-propositions, as $r$ sub-propositions, we get $u(x,y,..) \sim - r \lambda w ln(w)$, and so, from Eq.(\ref{eq:ignorance_notpondered}), we obtain

\begin{equation}\label{eq:generalagranger}
H[p_1,..,p_n]= \lambda(u) \left(\sum_{i=1}^n p_i - 1 \right) -   \sum_{i=1}^n  \lambda(r) (p_i ln(p_i) )+ \lambda[1]        
\end{equation}

where $[1]$ condense all the constants.\sk 

Commentaries about $\lambda(r)$ : As we said before, we used the sub-propositions to get the equation we need to solve. In our case, we "knew" that it exists $r$ sub-propositions, but someone may have known that only $r-1$ sub-propositions in the same case, thus leading to a factor $r-1$ instead of $r$. We could "cure" this reasoning saying that \textit{a priori} we do not know the $r$ sub-propositions, except that a proposition $A_w$ has at least two sub-propositions which are a sub-proposition and its contrary ($A_w = A + \bar{A}$) of probability $x$ and $y$ such that $x+y=1$. We could say that $A_w  = A$ of probability $x = 1$, and so $r=1$. However, a concern comes from that $A_j$, $j\neq 1$ constitute $\bar{A}_1$, so it would lead to a mix between the probabilities. 

However, just saying that there is one sub-proposition which is the proposition (of probability $x=1$), leads simply to $r=1$ in general (but also $ln(x) =1$ ...) , leading to the solution 

\begin{equation}\label{eq:rescalHlambda}	
H[p_1,..,p_n]= \lambda(u) \left[ \left(\sum_{i=1}^n p_i - 1 \right) -   \sum_{i=1}^n  (p_i ln(p_i) ) + [1] \right]       		
\end{equation}		 

\begin{itemize}
	\item We can always rescale $H[p]$ and deal with $h[p]$ such that $H[p] = \lambda h[p]$ as $\lambda$ is an arbitrary choice and the ignorance has to be the same for every individu with same knowledge on the situation : this expression has therefore to be invariant as such. 
	\item We can also take care of the constants in $[1]$ by always expressing information in terms of $H[p]- [1]$ , or ignorance relatively to maximum/minimum ignorance as $H[p]-H_m[p_m]$ for instance.  
\end{itemize}

When constraints are applied, the ignorance, also known as the \textit{information entropy} would therefore correspond to 
\begin{equation}\label{eq:finalexpressionE}
h[p_1,..,p_n]  \equiv  - \sum_{w=p_i} w \times ln(w)
\end{equation} 
as expected.

\subsubsection{cases where $b\geq2$}

As said previously, in these cases, we would have, roughly speaking,
\begin{equation}
\alpha(b,p) = \lambda(u) \left(\sum_{i=1}^n p_i -1 \right)^{b-1} \rightarrow \,\,\, [0]
\end{equation}
when applying the constraint. Moreover, $I_1$ will have the general form
\begin{equation}
I_1 = \sum_{v = 1}^{b} \gamma_1(v,c) \times (cx)^v
\end{equation}
where $\gamma_1(v)$ are numerical coefficients obtained after integrating ($b=2$, $\gamma_1(1) = -1$ and $\gamma_1(2) = -\frac{1}{2c}$). And for $I_2$, we obtain
\begin{equation}
I_2 =\sum_{v = 1}^{b-1} \gamma_2(v) \times (cx)^v - ln(x)  
\end{equation}
The solution can thus be expressed as 
\begin{equation}
u(x,y) = \beta\left(\dfrac{y}{x}\right) + [0]_{\lambda} + \sum_{v=1}^{b} \Gamma(v,c) (cx)^b + \dfrac{w\mu p}{b} ln x
\end{equation}
which becomes when applying the constraints $(cx-1)$ and restoring the symmetry
\begin{equation}
u(x,y) =  [0]_{\lambda} +[1]_{\mu} + \dfrac{w\mu p}{b} (ln x +lny )	
\end{equation}
or in general, doing the same simplifications as the case $b=1$, 
\begin{equation}
u(x,y) = [0]_{\lambda} +[1]_{\lambda,\mu} +  \dfrac{p\mu r}{b} \times w ln w	
\end{equation}
Commentaries : from this last equation, we could say that 

\begin{itemize}
	\item we can also rescale this expression in order to absorb the $[0]_{\lambda} +[1]_{\lambda,\mu}$ terms, and considering $\lambda=\dfrac{\mu}{b}$ we could also rescale as in Eq.(\ref{eq:rescalHlambda}), 
	\item $\lambda$ does not play a role at all, except to add constants via the $I_1$ term. Instead, it's really $\mu$ : $b=m$ and $\lambda = \mu$ seem to be irrelevant in the final expression, 
	\item $p=1$ is important, that is the constraint we add "outside", artificially. The constraint "inside", with $q$, which would make more sens in our opinion as it represents the sub-nodes, makes no effect (except for adding a constant) as in the first approach in this case. 
	
	\item we need to take $\bar{\mu} = -\mu$, or the constraint to be as $\mu(1-\sum_i v_i)$, in order for the ignorance to behave correctly, 
	
	\item and then, with these modifications, in these cases too, we obtain the expected expression for the ignorance to correspond to the \textit{information entropy}, for any value of $b$ (but in a less appealing way).

\end{itemize}

\section{Discussion and comment}
 
 Going back Eq.(\ref{eq:rescalHlambda}) 
concerns may raise about this expression where, constraints as $\lambda(u))$ lost its purpose if we express $H[p_1,..,p_n]$ as such. Indeed, the aim of $\lambda$ was here to take into account the fact that, normally,
\begin{equation}
 \dfrac{\partial}{\partial \lambda }H[p_1,..,p_n] = \sum_{i=1}^n p_i - 1 = 0   \,\,\, \Leftrightarrow \,\,\,\sum_{i=1}^n p_i = 1
\end{equation} 
which would not be the case here. 

One would rather assign different Lagrange multiplier such that, if we keep the general form similar as the one in Eq.(\ref{eq:generalagranger}), 
\begin{equation} 
H[p_1,..,p_n]= \lambda(u) \left(\sum_{i=1}^n p_i - 1 \right) +  \sum_{i=1}^n  \lambda(i) ( - p_i ln(p_i) + [1] )      
\end{equation}
one would obtain the following \textit{Equations of Motion}
\begin{eqnarray}
  \dfrac{\partial H}{\partial \lambda } &=& \sum_{i=1}^n p_i - 1 = 0   \,\,\, \Leftrightarrow \,\,\,\sum_{i=1}^n p_i = 1  \label{eq:constraintlamndasommeA}\\
 \dfrac{\partial H}{\partial \lambda(i) } &=& - p_i ln(p_i) + [1] = 0  \,\,\, \Leftrightarrow \,\,\, p_i ln(p_i) = value \\
\dfrac{\partial H}{\partial p_i  } &=& \lambda(u) + \lambda(i)(-ln(p_i) -1 ) = 0 \\
& \Leftrightarrow & p_i = exp\left[ \dfrac{\lambda(u) }{\lambda(i) -1 }\right]
\end{eqnarray}
In the case where $\lambda(u) = \lambda$ and $\lambda(i) = 1$, with the help of Eq.(\ref{eq:constraintlamndasommeA}), one would have 
\begin{eqnarray}
&&\sum_{i=1}^n p_i = 1 \Leftrightarrow   \sum_{i=1}^n (e^{\lambda-1}) = (e^{\lambda-1})  \times n  = 1 \\
&\Leftrightarrow & p_i = e^{\lambda-1} = \dfrac{1}{n}
\end{eqnarray}
which is, of course, the case of equiprobability where we only know only few things about $p_i$. In the case where for instance we know that $p_2 = 2 p_1$, one would be able to deal this situation by looking at the different $\lambda(i)$.

\section{Conclusion}

\begin{enumerate}

	\item  We have included constraints not as $\lambda \sum_i p_i$ as done for instance in \cite{Jaynesbook}, but as $\lambda(\sum_i p_i-1)$. This allows us to define what we call \textit{Ignorance $H$}, where 
	\begin{equation}
	H = H_{knowns} + H_{unknowns}
	\end{equation}
	where $H_{knowns} $ encodes the ignorance due to the constraints, therefore of zero ignorance. 
	
	\item In the first approach, we dealt with a quasi-known expression of the expression, \textit{i.e.} with the factor $p_i$ in front of $H_i$. In this case, it was like maximizing/minimizing the expected value of "local" sub-ignorance (at each branch of $p_i$) but leading to a final expression not really convincing as the logarithm term has to vanish when the constraints are applied. This was due, in our opinion, to the ill way of defining what happens at each sub-node such that,roughly speaking, $p_i H_i[p_i] \rightarrow p_i \times \dfrac{v_j}{p_i} H_i \left[ \dfrac{v_j}{p_i}\right]$. But we may have set it wrong and a more coherent way is possible.
	
	\item However, we found  way to cure this, starting from even before, not knowing at all the expression for the ignorance but just that it has also to apply in the same way at each node. Then we were able to get the expected expression for the Shannon entropy, but still with some interrogations linked to the same ones in the first approach.
	
	\item Mathematically, we have started from A but included sub-nodes as B in order to implement the fact that it has to be similar at each node. This helped us to obtain the correct expression for the differential expressions with the differentiation of the $\dfrac{1}{p_i}$ factors, leading to an expression in $w\,\, ln(w)$ primitive of $ln(w)+1$ and so the role of the exponential. 
	
	\item Moreover, we have also seen (at least partially) that the expression of the ignorance was somehow independent of the power taken for the constraints. In fact, the simplest case of power 1 seems in our opinion even better as we were able to obtain Eq.(\ref{eq:finalexpressionE}) in a coherent way, the higher power needing some adjustments. 
	
	\item In this way, the \textit{Maximization Entropy Principle} makes naturally sense as it is just the procedure to minimize our ignorance. It helped us to derive first the expression of the ignorance one has to obtain in order to be coherent, and secondly, knowing the expression but not the probabilities inside, to obtain these probabilities as usual and shown for instance in \cite{Jaynesbook}.
		
	\item Regarding the Lagrange multiplier, we were able to incorporate their subjectivity in an invariant way as the final expression of the Ignorance has to be the same whatever the choice of the multipliers. However, due to the presence of constants $[1]$, it would be better to express any quantity with respect to a reference value (as for temperature), that is, using $H[p]-H_m[p_m]$ for instance, in order to keep only the meaningful parts of the ignorance.	
  
  \item It is worth mentionning again the notion of 'surprise' $S(p)$ function of the probabilities \cite{bookRoss} and whose construction is similar as what we were looking at. It is based on axioms such that 
  \begin{enumerate}
  \item $S(1) = 0$ : no surprise if we know the outcome, that is, the ignorance is null. 
  \item $S$ is a decreasing function of $p$ : \\ 
  if $p<q$, then $S(p)>S(q)$.
  \item $S(p)$ is a continuous function of $p$. 
  \item Consider two independents events $E$ and $F$, of respective probabilities $p$ and $q$. The surprise of the event $EF$ of probability $P(EF) = pq$ would fulfill the equation 
  \begin{equation} 
  S(pq) = S(p) + S(q),
  \end{equation}
   \textit{i.e.} surprises are additive.   
  \end{enumerate}
  The function which satisfies these axioms is 
  \begin{equation}
  S(p) = - c \times ln(p)
  \end{equation}
  and the entropy is defined as the expected amount of surprise \begin{equation}
  H(X) = - \sum_{i} p_i ln (p_i).
  \end{equation}
    
 These axioms are shared by both approaches, and ignorance and surprise can be seen as the same object but with two ways of doing (the resolution of the surprise is however way shorter than the one for the ignorance where open issues still remain) and thinking : in our opinion, the surprise deals with independent events in a more "drastic" but direct way than what we did, and the fourth axiom constraints directly the shape of the solution. In our second approach, we just assumed additivity and updates (related also to independents propositions) and the key was to find the expression which minimize the ignorance dealing with constraints. As a results, we saw that it was similar to look at the expected amount of surprise : the subtleties are of course minimalistics, and we can consider both to be the same, just the framework and the way of thinking appear to be not present some differences.

\end{enumerate} 

To summarize : \sk 
Having knowledge on what we should have expected, we were biased but this helped us to start from zero and look at the situation from another perspective : having some notions about constraints and variational problems, reading the nice construction of the theory \cite{Jaynesbook} and on the maxmization entropy principle, gave us thoughts about including constraints on the probability in such a way that it could make sens. \sk 
As a consequence, we have defined general what we call "ignorance" and the procedure was "only" to try to minimize it (at least) and see if we could get back the correct expression for Shannon entropy: this is just the application of the maxmization entropy principle which appears naturally in this framework.\sk
To conclude, an extension of this work, at least in the way it has been done, may be helpful for instance in decision theory where one would define a quantity like the average risk, and try to minimize it as done here. This, however, will be kept for further researches. 

		\section{Acknowledgments}
	
	The author would like to express his deepest gratitude to Abhay, Aurelien, Martin, .. for time and space spend together. Thanks also to L\^e Nguy\^en Hoang for its pedagogical work which leads to look deeper to the Bayesian approach, David Aspnes and Will Perkins for discussions and pointing out the notion of surprise. Wolframalpha was used to check the calculations, and Geogebra to plot figures using tikz in LateX.
	


\end{document}